\documentclass{aa}
\usepackage{longtable}
\usepackage{multirow}
\usepackage{graphicx,natbib,txfonts,color}
\bibpunct{(}{)}{;}{a}{}{,}

\begin{document}

\title{The lead discrepancy in intrinsically $s$-process enriched post-AGB stars in the Magellanic Clouds.
\thanks{based on observations collected with the Very Large Telescope
  at the ESO Paranal Observatory (Chili) of programme number 084.D-0932 and 088.D-0433.} }

\author{K. De Smedt\inst{1}
\and H. Van Winckel\inst{1}
\and D. Kamath\inst{1}
\and A. I. Karakas\inst{2}
\and L. Siess\inst{3}
\and S. Goriely\inst{3}
\and P. Wood\inst{3}
}

\offprints{K. De Smedt, kenneth.desmedt@ster.kuleuven.be}

\institute{ Instituut voor Sterrenkunde, K.U.Leuven, Celestijnenlaan 200B,
B-3001 Leuven, Belgium
\and  Research School of Astronomy and Astrophysics, Mount Stromlo Observatory, 
Weston Creek ACT 2611, Australia
\and Institut d'Astronomie et d'Astrophysique, Universit\'{e} Libre de Bruxelles, ULB, CP 226, 1050 Brussels, Belgium
}

\date{Received  / Accepted}

\authorrunning{K. De Smedt et al.}
\titlerunning{Lead discrepancy of $s$-process enriched Magellanic Cloud stars}

\abstract
{ Our understanding of the $s$-process
  nucleosynthesis in asymptotic giant branch (AGB) stars is
  incomplete. AGB models predict, for example, large overabundances of
  lead (Pb) compared to other $s$-process 
elements in metal-poor low-mass AGB stars. This is indeed observed in
some extrinsically enhanced metal-poor stars, but not in all.
An extensive study of intrinsically $s$-process enriched objects is 
essential for improving our knowledge of the AGB third dredge-up and associated $s$-process nucleosynthesis.}
{We compare the spectral abundance analysis of the SMC post-AGB star J004441.04-732136.4 with state-of-the-art AGB model 
predictions with a main focus on Pb. The low S/N in the Pb line region made the result of our previous 
study inconclusive. We acquired additional data covering the region of the 
strongest Pb line.}
{By carefully complementing re-reduced previous data, with newly acquired UVES optical spectra, we improve the S/N 
of the spectrum around the strongest Pb line. Therefore, an upper limit for the Pb abundance is estimated from a merged weighted mean spectrum 
using synthetic spectral modeling. We then compare the abundance results from the combined spectra to predictions of tailored AGB evolutionary 
models from two independent evolution codes. In addition, we determine
upper limits for Pb abundances for three previously studied LMC post-AGB objects. }
{Although theoretical
predictions for J004441.04-732136.4 match the $s$-process distribution
up to tungsten (W), the predicted very high Pb abundance is clearly not detected. The three additional LMC post-AGB stars show a similar lack 
of a very high Pb abundance.}
{ From our study, we conclude that none of these low-mass, low-metallicity
  post-AGB stars of the LMC and SMC are strong Pb producers. This conflicts with current theoretical predictions.}


\maketitle


\section{Introduction}\label{sect:intro}

During the asymptotic giant branch (AGB), stars undergo thermal pulses
that may be followed by third dredge-ups during which freshly synthesized products are brought into the envelope, among them the carbon and $s$-process elements.
Theoretical and
observational evidence shows that the main neutron source 
in low-mass AGB stars (1-3 M$_{\odot}$) is the $^{13}$C($\alpha$,n)$^{16}$O reaction
\citep{straniero95,gallino98,abia02}. 
It is widely accepted that the
$^{13}$C pocket originates in a region where protons from the H-rich convective envelope are mixed into the He-rich intershell after a thermal pulse.
However, the physical
mechanisms of both third dredge-up and partial mixing in the intershell remain poorly 
understood.\newline 
The $s$-process nucleosynthesis is predicted to depend strongly
on metallicity while the $^{12}$C component, hence the $^{13}$C
neutron source, in the intershell are of primary origin. 
In metal-deficient environments ([Fe/H] $<$ -1), in proportion, more neutrons are available for each iron seed resulting in larger overabundances of heavy $s$-process elements with respect to lighter ones ([hs/ls]). 
However, to date no clear relation has been
observed between [hs/ls] and metallicity (see e.g. Fig 10 in
\cite{vanaarle11}) in post-AGB stars. The end product of the $s$-process nucleosynthesis chain is the double magic lead $^{208}$Pb isotope, which is predicted to have large overabundances with respect to other s-elements in metal-poor conditions
(see \cite{gallino98},\cite{goriely00} and \cite[][and references therein]{lugaro12}). The detection of metal-deficient objects with strong Pb enhancement
by e.g. \cite{vaneck01,vaneck03} and \cite{behara10} confirm these predictions. However, some 
low-metallicity extrinsically $s$-process enriched objects (i.e. binaries) were found without strong Pb overabundance 
\citep{aoki01,vaneck03}. These are just a few illustrations that
  show that our understanding of the $s$-process nucleosynthesis is
  limited. Additional systematic observations are required to
  deduce $s$-process distributions, including Pb abundances, to improve
  our understanding of the third dredge-up and
AGB nucleosynthesis mechanisms \citep[e.g.][and references therein]{herwig05}.\newline 
In \cite{desmedt12} (paper I), we performed an extensive spectral abundance analysis of the Small Magellanic Cloud (SMC)
post-AGB object \object{J004441.04-732136.4} (hereafter abbreviated 
to J004441). As a 21 $\mu$m source \citep{volk11}, J004441 shows very strong $s$-process enhancement combined with a low metallicity of [Fe/H] = 
-1.34 $\pm$ 0.32 and a modest C/O ratio of 1.9 $\pm$ 0.7. 
A luminosity of 7600 $\pm$ 200 L$_{\odot}$ was derived with the known distance to the SMC. Comparison of the position of J00441 in the HR diagram to evolutionary post-AGB tracks of 
\cite{vassiliadis94} resulted in an estimated initial mass of approximately 1.3 M$_{\odot}$. Based upon 
the initial mass and metallicity, different theoretical models were calculated with two independent stellar evolution 
codes, namely the Mount-Stromlo Evolutionary code \citep[][and references therein]{karakas10a} and the STAREVOL
code \citep[e.g.][and references therein]{siess07}. The theoretical predictions match the $s$-process distribution but
fail to reproduce the high overabundances and the modest C/O ratio. All models predict strong Pb overabundances but unfortunately, 
the strongest spectral Pb line at 4057.807 $\AA$ was only
covered by a very low S/N spectrum, preventing an accurate 
Pb abundance determination with the data of Paper I. We therefore reanalysed the Pb abundance of J004441 using a newly obtained UVES spectrum which is
described in this letter.\newline 
In addition, we determine upper limits for the Pb abundances for three Large Magellanic Cloud (LMC) post-AGB objects. Accurate spectral abundance studies of these objects have been conducted by \cite{vanaarle11}, but Pb was not included in their analysis. Comparison of the different Pb enhancements will provide insight into the intrinsic enrichment of metal-poor, low-mass AGB objects. \newline
In the following section, we report on the new data and Section~3
describes our analysis and abundance determination. We compare the
results of J004441 with model predictions in Section~4, followed by
analyses of the Pb abundances for the LMC objects in Section~5. We end with a brief discussion.\newline

\section{Observations and data reduction}\label{sect:data}
We use high-resolution spectra obtained with the UVES spectrograph \citep{dekker00_2} mounted on the VLT. 
Apart from the UVES spectra used in Paper I, we obtained a new UVES spectrum 
with an exposure time of 1500 seconds. The dichroic beam splitter was used, which provided a wavelength 
coverage from approximately 3760 to 4985 $\AA$ for the blue arm and 6705 to 8513 $\AA$ and 8663 to 10420 $\AA$ for the lower and 
upper parts of the mosaic CCD chip, respectively. This results in a wavelength range overlap with the Paper I spectra of
3760 to 4530 $\AA$, 4780 to 4985 $\AA$, and 6705 to 6810 $\AA$.\newline
Reduction of the UVES data is performed using the Reflex environment. In an attempt to optimize the S/N of the blue spectra, 
the new blue spectrum is reduced using different settings for the slit length
and wavelength bin size in the reduction pipeline. Then the S/N of the different spectra are determined by computing the standard
deviation obtained in different continuum regions in the wavelength range from 4800 $\AA$ to 
4950 $\AA$. This wavelength range is present in both the old and new spectra. For the new data, the best S/N was found for the standard 
parameter settings in the pipeline. Based upon the S/N results for the new data, the spectra of Paper I are also reduced again using the 
standard parameter settings in the pipeline.\newline
The different blue spectra are then merged into a weighted mean spectrum. The weights are based upon the S/N quality. A small spectral window
of 50 $\AA$ around the strongest 4057.807 $\AA$ Pb line is then normalized by fitting a fifth order polynomial through interactively defined continuum points.
The final S/N of this combined spectrum is $\sim$ 25 in the region of the Pb line.\newline
The velocity correction for the new data is exactly the same as for the previous data hence a heliocentric radial velocity of 148 $\pm$ 3 km/s is applied. 
This constant velocity suggests J004441 to be either a single star or a binary in a wide orbit.
 
\section{Spectral analysis of Pb in J004441}\label{sect:Pbabun}
\begin{table}[tb!]
\caption{\label{table:atmos} Determined atmospheric parameters of J004441 in Paper I and three LMC objects of \cite{vanaarle11}.}
\begin{center}
\begin{tabular}{lcccc} \hline\hline
           & J004441 & J050632 & J052043 & J053250 \\ \hline
$T_{eff}$ (K) & $6250 \pm 250$ & 6750 $\pm$ 250 & 5750 $\pm$ 250 & 5500 $\pm$ 250 \\
log g (dex) & $0.5 \pm 0.5$ & $0.5 \pm 0.5$ & $0.0 \pm 0.5$ & $0.0 \pm 0.5$\\
$\xi_t$ (km/s) & $3.5 \pm 0.5$ & $3.0 \pm 0.5$ & $3.0 \pm 0.5$ & $3.0 \pm 0.5$ \\
$\left[\textrm{Fe/H}\right]$ (dex) & -1.34 $\pm$ 0.32 & -1.22 $\pm$ 0.18 & -1.15 $\pm$ 0.20 & -1.22 $\pm$ 0.19 \\
L/L$_{\odot}$ & $7600 \pm 200$ & $5400 \pm 700$ & $8700 \pm 1000$ & $6500 \pm 1000$ \\
\hline
\end{tabular}
\end{center}
\end{table}

\begin{figure}[t!]
\resizebox{.97\hsize}{!}{\includegraphics{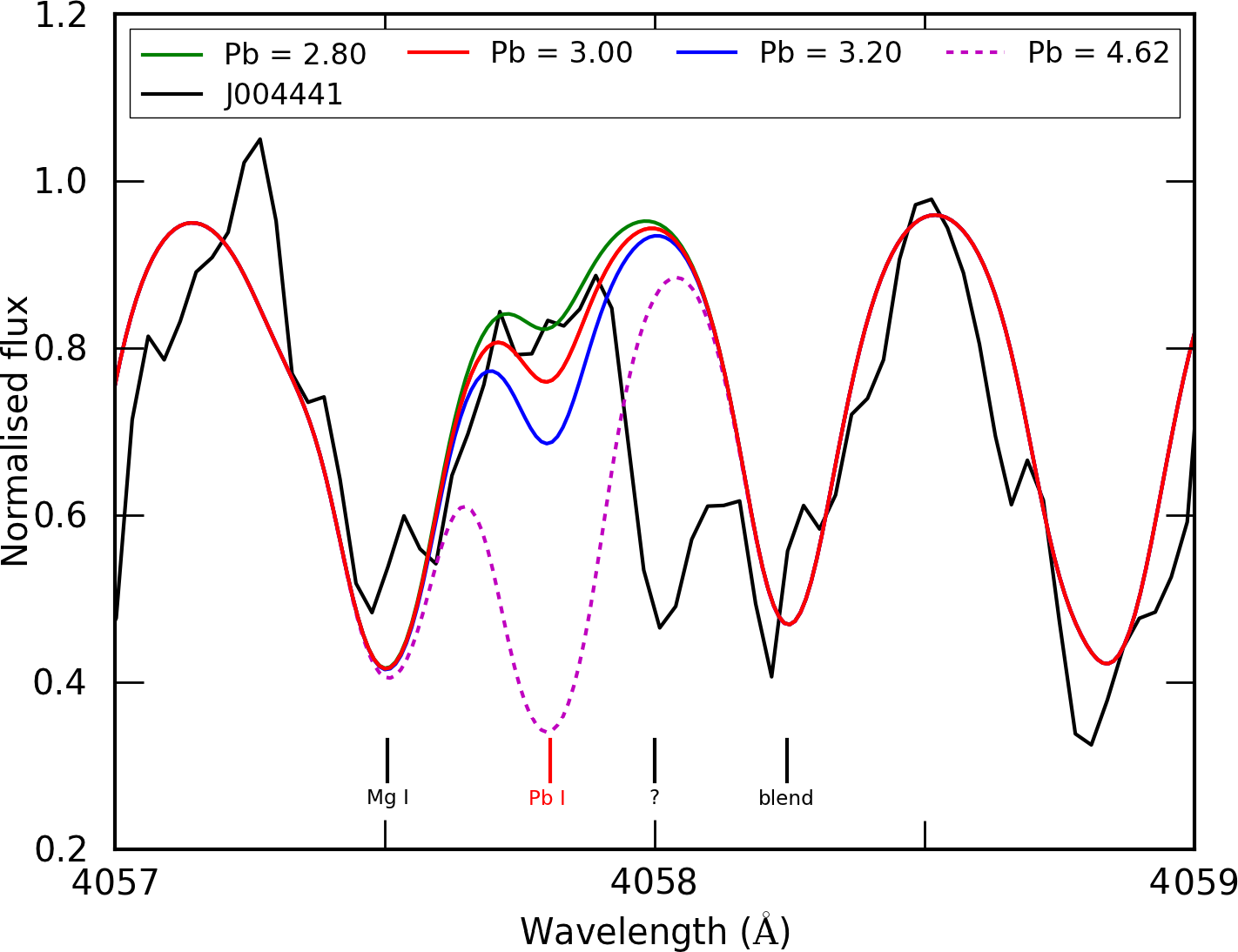}}
\caption{Upper limit Pb abundance determination of J004441 using
  spectrum synthesis. The black spectrum is J004441 while
coloured spectra represent synthetic spectra with different Pb abundances. Full coloured lines are used for the upper limit Pb abundance determination,
the dashed magenta line represents the line when assuming an Pb
abundance as predicted by the best fitting AGB model (see Sect. 4). The question mark indicates the position of an unidentified
spectral line.}\label{fig:fit}
\end{figure}

\begin{figure}[tb!]
\resizebox{.97\hsize}{!}{\includegraphics{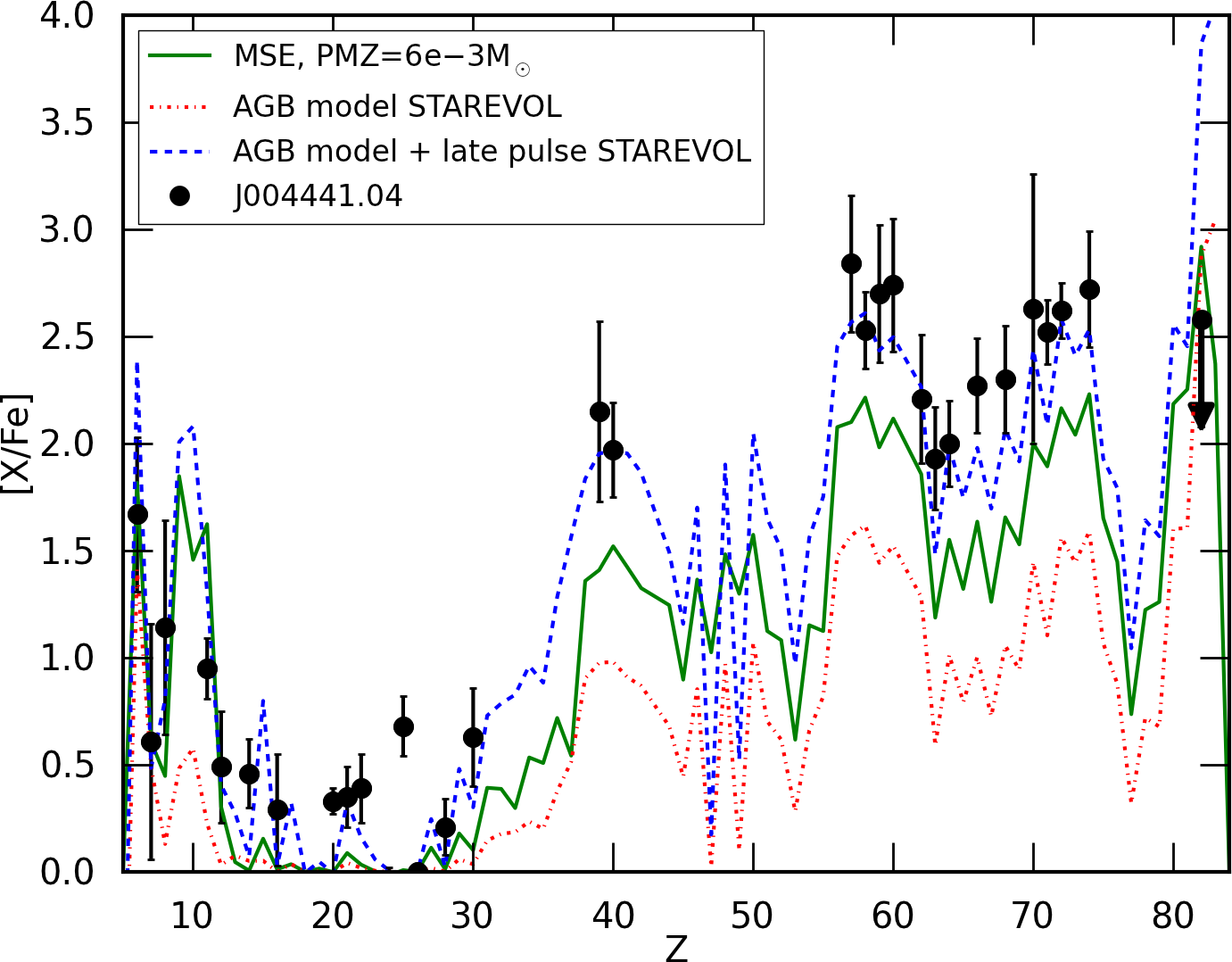}}
\caption{Comparison of the J004441 abundance ratios with model predictions. Black dots represent the [X/Fe] results of J004441,
the green line represents the Mount Stromlo (MSE) model,
red and blue dashed lines show STAREVOL code predictions with the latter including
a late thermal pulse with a deep dredge-up when dilution is small. The [Pb/Fe] result is indicated with a down arrow, representing an upper limit.}\label{fig:Pbmodel}
\end{figure}

For clarity, Table \ref{table:atmos} shows the atmospheric parameters
of J004441 (determined in Paper I), together with the derived parameters of three
LMC objects from \cite{vanaarle11}, which is discussed
below. We assume [Fe/H] as an accurate value of the mean metallicity of the stars. For all objects, 
we redid neither the atmospheric parameter
determination nor the abundance analysis of already determined
elements. \newline The merged spectrum quality is sufficient to
determine an upper limit of the Pb abundance using the region around the 4057.807 $\AA$ line. We used an LTE
Kurucz-Castelli atmosphere model \citep{castelli04} with the
atmospheric parameters shown in Table \ref{table:atmos}, combined
with the LTE abundance calculation program MOOG \citep{sneden73} and
VALD linelists \citep{vald}. The studied elemental abundances in Paper
I, together with an overall metallicity [M/H] = -1.34 dex, are included
in the spectral synthesis fit of the Pb line. Possible non-LTE effects
are not taken into account.\newline 
Figure \ref{fig:fit} displays the
spectral synthesis fits to the Pb line in J004441. The coloured spectra represent different abundance values for
Pb with all other elemental abundances constant. The upper
limit for the Pb abundance is derived by comparison with the full coloured
lines.  It is difficult to determine the continuum level accurately. The
fit of the Mg I line to the left of the Pb line and the blended
line at 4058.2 $\AA$ are used as estimates of the continuum
position.  The question mark in Fig.\ref{fig:fit}, 
indicates the wavelength position of a spectral line which remains
unidentified despite our efforts. Manually increasing the input abundance of all available
elements in the region of this line did not change the synthetic
spectrum, so we conclude that this line is not present in the
VALD linelist. In strongly $s$-process enhanced photospheres, the mere
line identification is often problematic \citep[e.g.][]{reyniers02}.
Considering the strong $s$-process enrichment of J004441, this line
is probably an unidentified $s$-process spectral line. The identification of
this spectral line is essential for constraining a more accurate Pb
abundance, and it allows us to set an upper limit for the Pb
abundance around log$_{\epsilon}$ (Pb) = 3.0 (log$_{\epsilon}$ (H) =
12.0).\newline To test the abundance fit, we replaced the VALD
Pb line in the VALD linelist by the analogue Pb line from the NIST
Atomic Spectra Database \citep{nist_2} and the Pb line analogue from the
Pb line list used in \cite{vaneck03}.  The oscillator strengths (log
gf) for the VALD, NIST, and Van Eck linelists for the Pb line are
-0.17 dex, -0.18 dex and -0.22 dex, respectively, while the excitation
potential is 1.32 eV for all linelists. This results in a maximal
abundance difference of 0.05 dex, which is neglible with respect to the
uncertainty due to the continuum position. Also the isotopic Pb line
list of \cite{vaneck03} was fitted assuming solar isotope ratios and
obviously resulted in an even lower Pb abundance upper limit, which
seems improbable considering the Pb abundance discrepancy in
Sect. \ref{sect:model}.  We therefore adopt a Pb abundance of
log$_{\epsilon}$(Pb) = 3.0 for comparison with AGB models.\newline 
The nitrogen (N) abundance is also derived using spectral synthesis. In Paper I, no
clear N lines were detected in the optical part of the
spectra. However, the new spectra cover a wider wavelength range
towards the IR, displaying clear unblended N
lines. Based upon the analysis of five spectral lines, we find
log$_{\epsilon}$ (N) = 7.10 $\pm$ 0.32 where the error contains
model, line-to-line scatter, and non-LTE uncertainties. We apply a
non-LTE correction of 0.3 dex as found in \cite{lyubimkov11} to
similar objects and adopt this value as the non-LTE uncertainty.

\section{AGB chemical models}\label{sect:model}

We compare the newly derived abundance results with model predictions
from two independent stellar evolution codes of Paper I. Both codes
calculated models for a 1.3 M$_{\odot}$ star of [Fe/H] = -1.4 for which the calculated 
AGB tip luminosities are somewhat higher than the observed luminosity. The luminosities can be 
fine-tuned by increasing the mass-loss rate, but this should not alter the
predicted abundance profiles.\newline
The observed abundances compared to model predictions are shown in
Fig. \ref{fig:Pbmodel} where different coloured lines represent
different model predictions. The derived [Pb/Fe] upper limit of J004441
is indicated with a down arrow. The Mount-Stromlo
Evolutionary (MSE) predictions \citep[][and references
therein]{karakas10a} are calculated for a metallicity of Z = 0.0006,
while both STAREVOL code predictions \citep[e.g.][and references
therein]{siess07} have a reference composition of Z = 0.0044 (see Paper I)
and are calculated using the same parameter values. The red STAREVOL model in 
includes a late thermal pulse combined with a deep dredge-up, 
resulting in substantial surface pollution due to the low convective envelope mass. 
A detailed description of the conclusions drawn from Fig. \ref{fig:Pbmodel} for
C, O, and s-elements except Pb, is given in Paper I and will not be
repeated in this letter. \newline
The dashed magenta line in Fig. \ref{fig:fit} is computed using
the [Pb/Fe] = 4.62 as predicted from the best fitting AGB 
model in Fig. \ref{fig:Pbmodel} and clearly shows that such a high Pb
abundance is incompatible with our spectrum. Figure \ref{fig:Pbmodel} also indicates
that the Pb overabundance is similar to that of the other heavy $s$-process peak 
elements (La,Ce and Nd around Z=60). The newly derived N abundance (Z =
7) fits the different model predictions well.\newline

\begin{table}[tb!]
\caption{\label{table:sfe} [s/Fe] results for specific s-elements of the studied objects.}
\begin{center}
\begin{tabular}{lcccc} \hline\hline
        & \textrm{[La/Fe]} & \textrm{[Ce/Fe]} & \textrm{[Nd/Fe]} & \textrm{[Pb/Fe]}  \\ \hline
J004441 & 2.84 $\pm$ 0.32  & 2.53 $\pm$ 0.18  & 2.74 $\pm$ 0.31  & $<$2.58 \\
J050632 & 1.48 $\pm$ 0.25  & 1.33 $\pm$ 0.23  & 1.18 $\pm$ 0.31  & $<$1.52 \\
J052043 & 1.85 $\pm$ 0.24  & 1.68 $\pm$ 0.20  & 1.92 $\pm$ 0.26  & $<$1.40 \\
J053250 & 2.03 $\pm$ 0.26  & 1.91 $\pm$ 0.20  & 2.02 $\pm$ 0.24  & $<$1.70 \\
\hline
\end{tabular}
\end{center}
\end{table}

\section{LMC objects}
\begin{figure}[tb!]
\resizebox{.97\hsize}{!}{\includegraphics{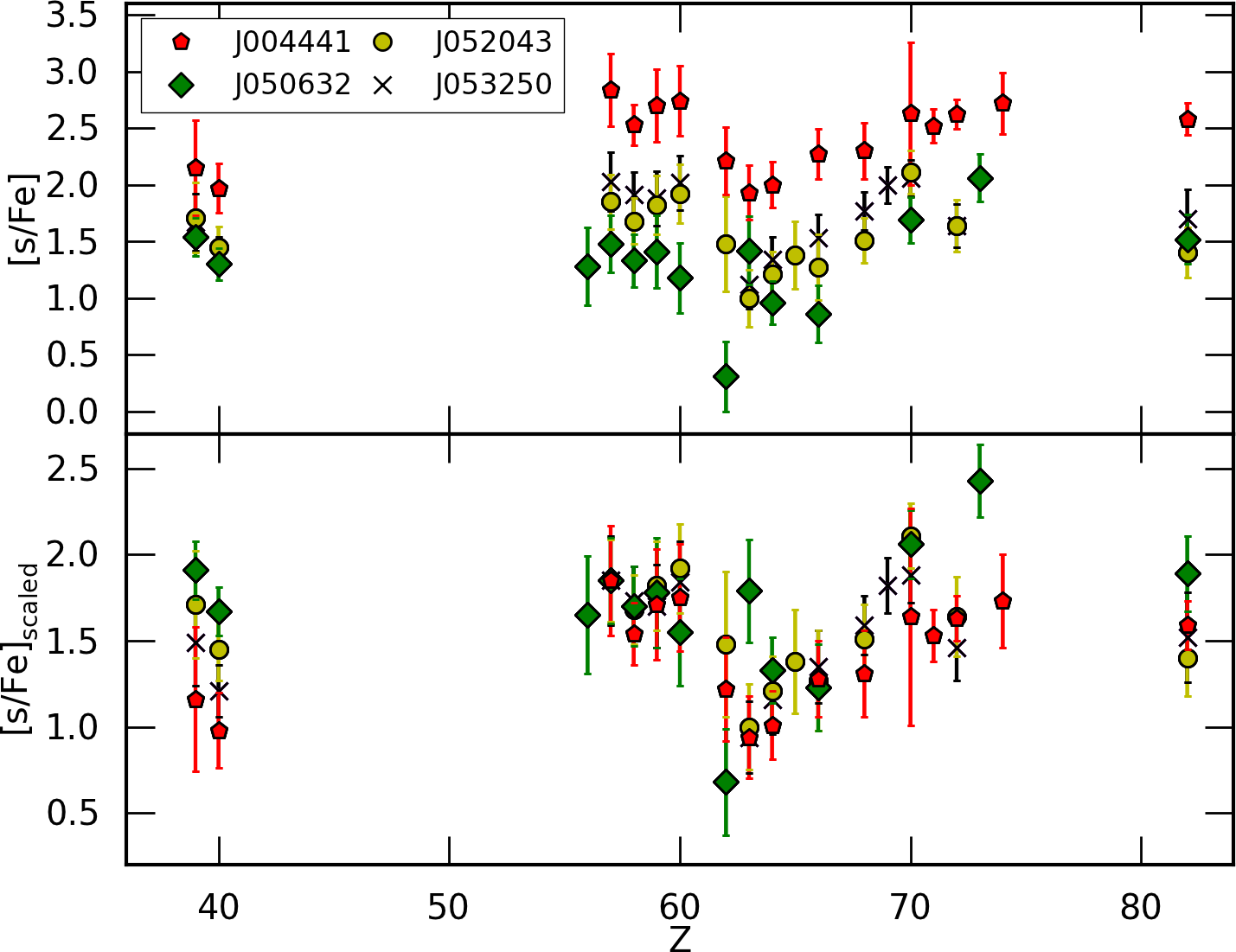}}
\caption{Upper panel: Overview of [s/Fe] for the different objects. The values at Z=82 represent the derived Pb abundance upper limits.
Lower panel: [s/Fe] for the different objects scaled to [La/Fe] at Z=57.}\label{fig:comp}
\end{figure}

To expand the sample of studied Pb abundances in $s$-process enriched
post-AGB stars, Pb abundance upper limits are derived for three LMC
objects studied by \cite{vanaarle11} (Paper II). These three objects
are \object{J050632.10-714229.8}, \object{J052043.86-692341.0}, and
\object{J053250.69-713925.8}, hereafter respectively abbreviated to
J050632, J052043, and J053250. For the Pb abundance analysis, we use
the normalized spectra of Paper II. For J052043, two studies were
conducted in Paper II of which we use the spectra observed in December
(J052043\_b in Paper II). Table \ref{table:atmos} shows the
atmospheric parameters of the three objects. All objects are found to
be metal-poor and to have initial masses below 1.5 M$_{\odot}$. In none of
the objects is the Pb line clearly detected.\newline Similar to the
study of J004441, spectral synthesis is used for the Pb abundance
determinations of the three LMC objects. Table \ref{table:sfe} lists 
the [s/Fe] determinations for Ba-peak elements, along with the Pb
abundance upper limits. The shown abundance results of the three LMC
objects are derived in Paper II, except for Pb. For none of the
objects is the [Pb/Fe] uncertainty larger than 0.08 dex due to model
uncertainties. The [s/Fe] results for all studied s-elements of the
different objects are shown in the upper panel of
Fig. \ref{fig:comp}. For a given metallicity, the
overabundances depend on the amount of dilution in the envelope
mass. Therefore, the lower panel of Fig. \ref{fig:comp} shows the
[s/Fe] results scaled to [La/Fe] of J052043 to show the $s$-process
distribution of the different stars independent of dilution. From
Table \ref{table:sfe} and Fig. \ref{fig:comp}, we conclude that for
J004441 and J050632, the Pb upper limit overabundance is similar to the
overabundances of the  Ba-peak
elements, while for J052043 and J053250, Pb is modestly
underabundant with respect to Ba-peak elements.

\section{Discussion and conclusion}\label{sect:disc}
Using additional spectra of J004441, an $s$-process rich post-AGB star
in the SMC, we obtained an upper limit of the Pb abundance.
With [Pb/Fe] $<$\,2.58,  Pb shows a similar or smaller
overabundance than the $s$-process elements of the Ba-peak.  However, this upper limit is much smaller than 
predicted by our specific AGB models. The latter were tuned to the
initial mass and metallicity of the object and focussed on a fit
through the obtained overabundances of the Zr- and Ba-peak elements.
This discrepancy is real and not due to the poor S/N of the
spectrum. The $s$-process predictions of the tuned AGB models, based
on successive thermal pulses with associated partial mixing
and dredge-ups are not able to reproduce the low Pb abundance.\newline
The [Pb/Fe] results for the three LMC objects are similar to J004441: there is no
predicted high Pb overabundance relative to the Ba-peak elements.
Although having a strong $s$-process enrichment,
  the four low-mass, metal-deficient post-AGB stars of the Magellanic clouds all show the absence of the theoretically predicted strong Pb
overabundances. 
This trend is similar to what is observed in some, but
not all, extrinsically enriched objects in
\cite{aoki01} and \cite{vaneck03}. 
Current  AGB models based on a $^{13}$C-pocket arising from diffusive
overshooting at the base of the convective envelope during the third
dredge-up have problems reproducing the observed low Pb
abundances in metal poor stars, the spread in [hs/ls] at a given
metallicity and/or the C/O ratios. This clearly indicates that some
physical ingredients are missing in the description of this process. To
improve the situation, we need to obtain full abundance patterns of many post-AGB stars in both
the LMC and SMC, so that we can identify systematics in the observed
patterns. Simultaneously, we need to explore alternative processes to
explain the derived abundance profiles. In this study, we used the diffusive approach to form
the $^{13}$C pocket, but the partial mixing of protons below the envelope
may not be diffusive in essence but rather advective. For instance, when adopting a new mixing
algorithm, \cite{straniero06} were able to generate a more massive $^{13}$C pocket and
consequently produce higher $s$-process surface enrichments in better
agreement with observations. Rotation is also a key physical ingredient
that may have a strong impact on the synthesis of $s$-elements. Depending
on the degree of shear, the $s$-process nucleosynthesis can be completely
inhibited (see e.g. \cite{langer99}, \cite{herwig03} and \cite{siess04}) or partially activated \citep{piersanti13}. Also the effect of internal gravity 
wave mixing \citep{denissenkov03} on the $s$-process nucleosynthesis 
still needs to be explored. \newline
Additional work, both theoretically assisted by direct hydrodynamical
simulations and observational with systematic abundance determinations
of a whole population of post-AGB stars with well constrained distances
and a spread in metallicity, is needed to better constrain the $s$-process
mechanisms.

\bibliographystyle{aa}
\bibliography{allreferences}

\begin{thebibliography}{29}
\expandafter\ifx\csname natexlab\endcsname\relax\def\natexlab#1{#1}\fi

\bibitem[{{Abia} {et~al.}(2002){Abia}, {Dom{\'{\i}}nguez}, {Gallino}, {Busso},
  {Masera}, {Straniero}, {de Laverny}, {Plez}, \& {Isern}}]{abia02}
{Abia}, C., {Dom{\'{\i}}nguez}, I., {Gallino}, R., {et~al.} 2002, \apj, 579,
  817

\bibitem[{{Aoki} {et~al.}(2001){Aoki}, {Ryan}, {Norris}, {Beers}, {Ando},
  {Iwamoto}, {Kajino}, {Mathews}, \& {Fujimoto}}]{aoki01}
{Aoki}, W., {Ryan}, S.~G., {Norris}, J.~E., {et~al.} 2001, \apj, 561, 346

\bibitem[{{Behara} {et~al.}(2010){Behara}, {Bonifacio}, {Ludwig}, {Sbordone},
  {Gonz{\'a}lez Hern{\'a}ndez}, \& {Caffau}}]{behara10}
{Behara}, N.~T., {Bonifacio}, P., {Ludwig}, H.-G., {et~al.} 2010, \aap, 513,
  A72

\bibitem[{{Castelli} \& {Kurucz}(2004)}]{castelli04}
{Castelli}, F. \& {Kurucz}, R.~L. 2004, ArXiv Astro-ph/0405087

\bibitem[{{De Smedt} {et~al.}(2012){De Smedt}, {Van Winckel}, {Karakas},
  {Siess}, {Goriely}, \& {Wood}}]{desmedt12}
{De Smedt}, K., {Van Winckel}, H., {Karakas}, A.~I., {et~al.} 2012, \aap, 541,
  A67

\bibitem[{{Dekker} {et~al.}(2000){Dekker}, {D'Odorico}, {Kaufer}, {Delabre}, \&
  {Kotzlowski}}]{dekker00_2}
{Dekker}, H., {D'Odorico}, S., {Kaufer}, A., {Delabre}, B., \& {Kotzlowski}, H.
  2000, in Society of Photo-Optical Instrumentation Engineers (SPIE) Conference
  Series, ed. M.~{Iye} \& A.~F. {Moorwood}, Vol. 4008, 534--545

\bibitem[{{Denissenkov} \& {Tout}(2003)}]{denissenkov03}
{Denissenkov}, P.~A. \& {Tout}, C.~A. 2003, \mnras, 340, 722

\bibitem[{{Gallino} {et~al.}(1998){Gallino}, {Arlandini}, {Busso}, {Lugaro},
  {Travaglio}, {Straniero}, {Chieffi}, \& {Limongi}}]{gallino98}
{Gallino}, R., {Arlandini}, C., {Busso}, M., {et~al.} 1998, \apj, 497, 388

\bibitem[{{Goriely} \& {Mowlavi}(2000)}]{goriely00}
{Goriely}, S. \& {Mowlavi}, N. 2000, \aap, 362, 599

\bibitem[{{Herwig}(2005)}]{herwig05}
{Herwig}, F. 2005, \araa, 43, 435

\bibitem[{{Herwig} {et~al.}(2003){Herwig}, {Langer}, \& {Lugaro}}]{herwig03}
{Herwig}, F., {Langer}, N., \& {Lugaro}, M. 2003, \apj, 593, 1056

\bibitem[{{Karakas}(2010)}]{karakas10a}
{Karakas}, A.~I. 2010, \mnras, 403, 1413

\bibitem[{Kramida {et~al.}(2013)Kramida, {Yu.~Ralchenko}, Reader, \& {and NIST
  ASD Team}}]{nist_2}
Kramida, A., {Yu.~Ralchenko}, Reader, J., \& {and NIST ASD Team}. 2013, {
  National Institute of Standards and Technology, Gaithersburg, MD.}

\bibitem[{{Kupka} {et~al.}(1999){Kupka}, {Piskunov}, {Ryabchikova}, {Stempels},
  \& {Weiss}}]{vald}
{Kupka}, F., {Piskunov}, N., {Ryabchikova}, T.~A., {Stempels}, H.~C., \&
  {Weiss}, W.~W. 1999, \aaps, 138, 119

\bibitem[{{Langer} {et~al.}(1999){Langer}, {Heger}, {Wellstein}, \&
  {Herwig}}]{langer99}
{Langer}, N., {Heger}, A., {Wellstein}, S., \& {Herwig}, F. 1999, \aap, 346,
  L37

\bibitem[{{Lugaro} {et~al.}(2012){Lugaro}, {Karakas}, {Stancliffe}, \&
  {Rijs}}]{lugaro12}
{Lugaro}, M., {Karakas}, A.~I., {Stancliffe}, R.~J., \& {Rijs}, C. 2012, \apj,
  747, 2

\bibitem[{{Lyubimkov} {et~al.}(2011){Lyubimkov}, {Lambert}, {Korotin},
  {Poklad}, {Rachkovskaya}, \& {Rostopchin}}]{lyubimkov11}
{Lyubimkov}, L.~S., {Lambert}, D.~L., {Korotin}, S.~A., {et~al.} 2011, \mnras,
  410, 1774

\bibitem[{{Piersanti} {et~al.}(2013){Piersanti}, {Cristallo}, \&
  {Straniero}}]{piersanti13}
{Piersanti}, L., {Cristallo}, S., \& {Straniero}, O. 2013, \apj, 774, 98

\bibitem[{{Reyniers} {et~al.}(2002){Reyniers}, {Van Winckel}, {Bi{\'e}mont}, \&
  {Quinet}}]{reyniers02}
{Reyniers}, M., {Van Winckel}, H., {Bi{\'e}mont}, E., \& {Quinet}, P. 2002,
  \aap, 395, L35

\bibitem[{{Siess}(2007)}]{siess07}
{Siess}, L. 2007, \aap, 476, 893

\bibitem[{{Siess} {et~al.}(2004){Siess}, {Goriely}, \& {Langer}}]{siess04}
{Siess}, L., {Goriely}, S., \& {Langer}, N. 2004, \aap, 415, 1089

\bibitem[{{Sneden}(1973)}]{sneden73}
{Sneden}, C.~A. 1973, PhD thesis, The Univerisity of Texas at Austin.

\bibitem[{{Straniero} {et~al.}(1995){Straniero}, {Gallino}, {Busso}, {Chiefei},
  {Raiteri}, {Limongi}, \& {Salaris}}]{straniero95}
{Straniero}, O., {Gallino}, R., {Busso}, M., {et~al.} 1995, \apjl, 440, L85

\bibitem[{{Straniero} {et~al.}(2006){Straniero}, {Gallino}, \&
  {Cristallo}}]{straniero06}
{Straniero}, O., {Gallino}, R., \& {Cristallo}, S. 2006, Nuclear Physics A,
  777, 311

\bibitem[{{van Aarle} {et~al.}(2011){van Aarle}, {van Winckel}, {Lloyd Evans},
  {Ueta}, {Wood}, \& {Ginsburg}}]{vanaarle11}
{van Aarle}, E., {van Winckel}, H., {Lloyd Evans}, T., {et~al.} 2011, \aap,
  530, A90

\bibitem[{{Van Eck} {et~al.}(2001){Van Eck}, {Goriely}, {Jorissen}, \&
  {Plez}}]{vaneck01}
{Van Eck}, S., {Goriely}, S., {Jorissen}, A., \& {Plez}, B. 2001, \nat, 412,
  793

\bibitem[{{Van Eck} {et~al.}(2003){Van Eck}, {Goriely}, {Jorissen}, \&
  {Plez}}]{vaneck03}
{Van Eck}, S., {Goriely}, S., {Jorissen}, A., \& {Plez}, B. 2003, \aap, 404,
  291

\bibitem[{{Vassiliadis} \& {Wood}(1994)}]{vassiliadis94}
{Vassiliadis}, E. \& {Wood}, P.~R. 1994, \apjs, 92, 125

\bibitem[{{Volk} {et~al.}(2011){Volk}, {Hrivnak}, {Matsuura}, {Bernard-Salas},
  {Szczerba}, {Sloan}, {Kraemer}, {van Loon}, {Kemper}, {Woods}, {Zijlstra},
  {Sahai}, {Meixner}, {Gordon}, {Gruendl}, {Tielens}, {Indebetouw}, \&
  {Marengo}}]{volk11}
{Volk}, K., {Hrivnak}, B.~J., {Matsuura}, M., {et~al.} 2011, \apj, 735, 127

\end{thebibliography}

\end{document}